# Stepwise quantum phonon pumping in plasmon-enhanced Raman scattering

*Tian Yang[*], Jing Long and Xiaodan Wang*

State Key Laboratory of Advanced Optical Communication Systems and Networks, Key Laboratory for Thin Film and Microfabrication of the Ministry of Education, UM-SJTU Joint Institute, Shanghai Jiao Tong University, Shanghai 200240, China.

[*]tianyang@sjtu.edu.cn

**Abstract**

Plasmon-enhanced Raman scattering (PERS) becomes nonlinear when phonon pumping and phonon-stimulated scattering come into play. It is fundamental to the understanding of PERS and its photobleaching behavior. By quantization of the molecular vibration coherent state into phonon number states, we theoretically predict a stepwise dependence of PERS intensity on laser power. Experimental evidence is presented by measuring a monolayer of malachite green isothiocyanate molecules sandwiched in individual gold nanosphere-plane antennas, under radially polarized laser excitation of sub-$\mu$W powers.

It is well known that ultrahigh enhancement factors (EF) of plasmon-enhanced Raman scattering (PERS), including surface enhanced Raman scattering (SERS) and tip-enhanced Raman scattering (TERS), can be obtained by confining the exciting photons and Stokes photons in nanoscale hotspots. Such high EF substrates and devices have been investigated extensively for identification of minimum chemical traces and even studies on single molecule structures, dynamics and chemistry.[1-11] However, it is also widely accepted that the enhancement mechanism and photobleaching effect of PERS have not been completely understood. It is not until year 2013 that the nonlinear dependence of EF on laser power was revealed and taken seriously,[6-8] which is not contained by the classical $|E^4|$ law.[12] Borrowing from cavity optomechanics, a theoretical work was published three years later to attribute the nonlinearity to the dynamic backaction of molecular vibrations on plasmons, or equivalently, phonon stimulated Raman scattering.[13] The model also attributes the surprising improvement of imaging resolution in recent TERS experiments to a fast increase of EF near the phonon lasing threshold.[6,7,14] More recently, a full quantum electrodynamics model was established based upon the same concept of phonon stimulated Raman scattering.[15] These breakthroughs shed light on a deeper understanding of PERS.

However, there is limited experimental evidence to support the new theory. The experimental nonlinear single molecule TERS behaviors in ref. 6 and 7 were attributed by the authors to photon stimulated Raman scattering (SRS) instead, which could be possible under high laser powers. The experimental SERS EF versus laser wavelength spectrum in ref. 16 doesn't clearly show a threshold turn-on behavior as claimed by the theoretical work. Neither has narrowing of Raman lines been observed due to complicating experimental factors.[13] Further, in a recent experiment reported by us, we noticed that a single-stage fitting of the nonlinear SERS data will extrapolate to a finite SERS signal under no laser excitation which is certainly impossible.[8] Therefore, a more convincing experiment which clearly shows the signature of molecular phonon pumping in a plasmonic hotspot is needed for both understanding and further investigation in this direction. Here we address this problem by theoretically predicting and then experimentally observing stepwise quantum phonon pumping in single SERS hotspots. The stepwise process includes saturation of transitions between certain molecular vibration eigenstates under low laser powers, and stimulated emission of phonons at higher laser powers. The discovered phenomena provide more insight to the understanding and utilization of PERS.

In the following, first we describe a theoretical model which discriminates between pumping of different phonon number states, which are molecular vibration eigenstates that are occupied by different numbers of phonons. Previous models have assumed a homogeneous Raman activity for all these states, which actually reduces the problem to a single-step spontaneous and stimulated emission of phonons scenario.[13,15] In our model, arising from quantization of molecular vibration, a more complicated PERS intensity *vs* laser power relation than previously perceived is shown, due to stepwise pumping of phonons. Then we show experimental results using resonant Raman molecules to confirm the theoretical predictions. The experiment scheme, which is gold nanosphere – atomically flat gold plane antenna under radially polarized (RP) excitation, ensures reproducible ultrahigh EFs in deterministic hotspots.[8] It enables us to observe similar stepwise PERS behaviors repeatedly under sub-$\mu$W laser powers. With much less than one plasmon in the antenna, complicating factors such as SRS and heating are ignored. At the end, we discuss the validity of our model and relevant future work, including significantly slower vibration damping in dry PERS hotspots and photobleaching by phonon pumping.

**Results**

**The SERS experiment considered in this paper.** Our SERS measurement experiment was introduced in ref. 8 in details. For completeness of description, here we repeat the closely relevant parts, including the figures. A schematic illustration of the experiment is shown in Figure 1a. A 60 nm gold nanosphere is on top of an atomically flat gold plane, and an RP He-Ne laser beam at 633 nm is focused by an objective with a numerical aperture (NA) of 0.9 to excite the nanosphere. A monolayer of malachite green isothiocyanate (MGITC) molecules is coated on the surface of the nanosphere. The nanosphere pairs with its mirror image to form a vertical optical antenna, which contains a ~9 nm$^2$ hotspot in its junction gap. Raman scattering off molecules in the hotspot is collected by the same objective. Figure 1b shows a typical localized surface plasmon resonance (LSPR) spectrum of an antenna. The laser wavelength and three strong Raman bands of MGITC at 1180, 1370 and 1618 cm$^{-1}$ are labeled. These are the Raman bands we pay attention to in this paper. Figure 1c shows a typical SERS spectrum. Under a laser power of 300 nW at sample, measurement of twenty antennas has shown ultrahigh and reproducible electromagnetic EFs for the three Raman bands, which are $10^{9.2\pm0.2}$, as reported in ref. 8.

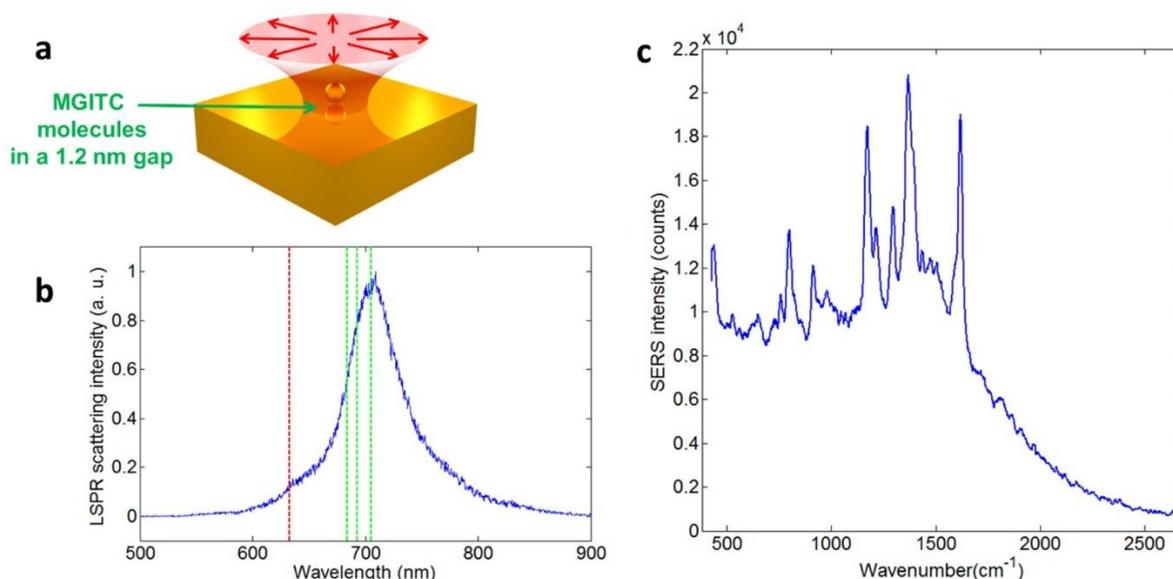

**Figure 1. LSPR and SERS of antennas coated with a monolayer of MGITC.**[8] (**a**) Illustration of the gold nanosphere-plane antenna under RP excitation. The mirror image of the nanosphere is also plotted. (**b**) The LSPR

spectrum of an antenna. The laser wavelength and three strong Raman bands in the SERS experiment are labeled as red and green lines, respectively. (**c**) The SERS spectrum of an antenna. The laser power at sample is 300 nW, and the integration time is 4 s.

**Theory: 1. Assumptions.** Our experiment allows the following assumptions to be made in order to simplify the theoretical model. (1) Under sub-$\mu$W laser power, the average number of plasmons and photons in the antenna is no more than 0.02, including the incident, the Stokes, plasmon decay to LSPR frequencies, lossy surface waves and fluorescence. A simple calculation of the average number of plasmons and photons was presented in ref. 8. Therefore, SRS effect is ignored, and the electromagnetic field will be treated classically, leading to a semi-classical treatment in our model. (2) As shown in Figure 1b, the laser is sitting at the shorter wavelength tail of LSPR, while the Stokes photons are close to the peak of LSPR. Therefore, anti-Stokes processes are much less enhanced by LSPR and will be ignored. (3) Since the Stokes shifts of the three Raman bands are large, the numbers of phonons in the corresponding molecular vibration modes are tiny at thermal equilibrium at room temperature, which is described by the Bose-Einstein Distribution. Therefore the thermal equilibrium phonons will be ignored since they have little effect on Stokes processes, and anti-Stokes processes are not considered as stated in Assumption (2). (4) The laser at 633 nm will be considered the only source of photons that drive the molecular vibrations. (5) The temperature effect will not be included, which will be briefly mentioned again when we discuss photobleaching.

**Theory: 2. Stokes transition between two neighboring molecular vibration eigenstates.** The influence of fluorescence will not be included in the main text since MGITC is known to be a non-fluorescent molecule, but will be discussed in Supplementary Information. The semi-classical differential PERS scattering cross section of a single molecule is given by ref. 12

$$\frac{d\sigma}{d\Omega} = \frac{\omega_S^4}{16\pi^2\varepsilon_0^2 c^4} |\tilde{E}_{Loc}(\omega_S) \cdot \hat{\alpha} \cdot \tilde{E}_{Loc}(\omega_L)|^2, \quad (1)$$

where $\sigma$ is PERS cross section, $\Omega$ is solid angle, $\omega_L$ and $\omega_S$ are the angular frequencies of the laser photons and Stokes photons respectively, $\varepsilon_0$ is dielectric constant of vacuum, and c is speed of light in vacuum. $\tilde{E}_{Loc}(\omega_L)$ is the local electric field in the hotspot normalized by that of the laser focal spot, therefore it is equal to local electric field enhancement at frequency $\omega_L$. $\tilde{E}_{Loc}(\omega_S)$ is local electric field enhancement at frequency $\omega_S$. $\hat{\alpha}$ is the Raman scattering polarizability tensor of a single molecule.

In equation (1), the $\hat{\alpha} \cdot \tilde{E}_{Loc}(\omega_L)$ term gives the dipole moment of the molecular vibration at the Stokes frequency. The $\tilde{E}_{Loc}(\omega_S)$ term is the Purcell effect which modifies this dipole's radiation. Hence the classical $|\tilde{E}_{Loc}^4|$ law for PERS EF is explicitly contained. The nonlinear PERS intensity versus laser power relation in previous models solely comes from the dependence of $\hat{\alpha}$ on laser power, as well as in our model.

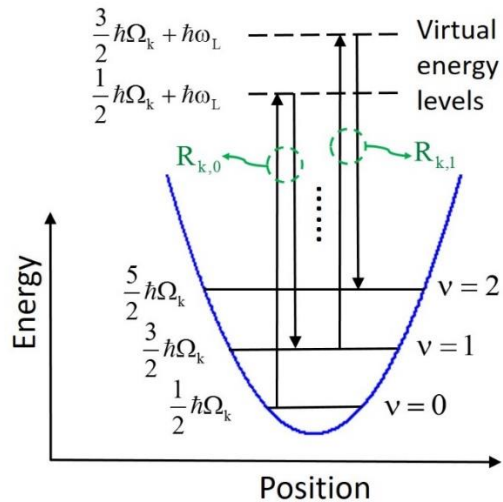

**Figure 2. Schematic illustration of the Stokes processes of a molecular vibration mode k.** The blue curve is the potential energy of the molecular vibration harmonic oscillator. Each phonon number state is labeled by a value of $v$, together with its energy. The Stokes processes $|v=0\rangle \to |v=1\rangle$ and $|v=1\rangle \to |v=2\rangle$ are plotted, together with their virtual energy levels and Raman activities.

Figure 2 is a schematic illustration of the Stokes processes. The molecular vibration of a normal mode, k, is treated as a harmonic oscillator with an effective mass $m_k$, a frequency $\Omega_k$ and a normal coordinate $Q_k$. The vibration eigenstates will be labeled as $|v\rangle$, where $v$ is the number of phonons. Each Stokes process increases the value of $v$ by 1. Quantum mechanically, the $|v\rangle \to |v+1\rangle$ Stokes process, that is, its $\hat{\alpha}$, is described by the following equations[12]

$$\hat{\alpha}_{k,v} = \langle v+1|\hat{Q}_k|v\rangle \hat{R}_{k,v} \ , \tag{2}$$

$$\hat{Q}_k = \sqrt{\hbar/2m_k\Omega_k}\left(\hat{b}_k + \hat{b}_k^+\right) \ , \tag{3}$$

$$\hat{R}_{k,v} = \left.\frac{\partial \hat{\alpha}_{k,v}}{\partial Q_k}\right|_{Q_k=0} , \tag{4}$$

where $\hat{Q}$ is the displacement position operator, $\hat{R}$ is the Raman activity tensor, and $\hat{b}$ and $\hat{b}^+$ are the annihilation and creation operators of the harmonic oscillator respectively. For simplicity, in the following, $\hat{R}$ and $\tilde{E}_{Loc}$ will be treated as scalars. Equations (1), (2) and (3) lead to a semi-classical expression for the differential PERS scattering cross section of the $|v\rangle \to |v+1\rangle$ Stokes process

$$\frac{d\sigma}{d\Omega} = \frac{\omega_S^4}{16\pi^2\varepsilon_0^2 c^4}\frac{\hbar}{2m_k\Omega_k}|\tilde{E}_{Loc}(\omega_S)|^2|\tilde{E}_{Loc}(\omega_L)|^2|R_{k,v}|^2(v+1) \ . \tag{5}$$

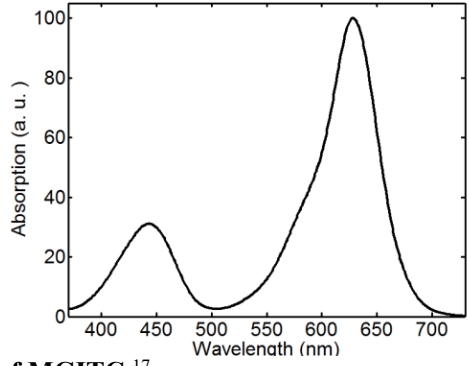

**Figure 3. Absorption spectrum of MGITC.**[17]

The fact that $R_{k,v}$ is dependent upon $v$ leads to a stepwise behavior which is different from previous models. The absorption spectrum of the probe Raman molecule in our experiments, MGITC, is shown in Figure 3.[17] The absorption peaks at 629 nm, which is very close to the laser wavelength. Therefore, the $|v=0\rangle \rightarrow |v=1\rangle$ Stokes process, as illustrated in Figure 2, involves a virtual state that is very close to one or a set of real vibrational states in an electronic excited state, and $R_{k,0}$ is strongly resonance enhanced. Although the $|v=1\rangle \rightarrow |v=2\rangle$ Stokes process may involve higher vibrational states of the same electronic excited state so that it is also resonance enhanced, the corresponding spatial overlap between the nuclei wavefunctions of the vibrational states in the electronic ground state and those in the electronic excited state is usually different from that of $|v=0\rangle \rightarrow |v=1\rangle$, according to the Franck–Condon principle energy diagram. For simplicity, in the numerical examples in the next section, we will take only three values of $R_k$, which are $R_{k,0}$, $R_{k,1}$, and $R_{k,2}$, and assume $R_{k,v} = R_{k,2}$ for all values of $v \geq 2$.

**Theory: 3. Quantum phonon pumping and stepwise PERS behavior.** According to equation (5), the phonons are pumped from $|v\rangle$ to $|v+1\rangle$ by the Stokes process at a speed of

$$\frac{1}{\hbar\omega_S}\frac{d\sigma}{d\Omega}\Omega_t I = \frac{1}{\hbar\omega_S}\frac{\omega_S^4}{16\pi^2\varepsilon_0^2 c^4}\frac{\hbar}{2m_k\Omega_k}|\tilde{E}_{Loc}(\omega_S)|^2|\tilde{E}_{Loc}(\omega_L)|^2|R_{k,\nu}|^2(\nu+1)\Omega_t I = (\nu+1)r_{k,\nu}I, \quad (6)$$

where $\Omega_t$ is the total solid angle of PERS, and I is laser intensity. All the dependences on the antenna, the molecule and the geometry of excitation optics are included in a transition rate coefficient $r_{k,\nu}$, which doesn't change with I, and which is proportional to $|R_{k,\nu}|^2$ for different values of $\nu$.

Let the average number of phonons in mode k to be $\bar{\nu}$, we have the following expression for a coherent state of vibration in mode k[18]

$$|\bar{\nu}\rangle = e^{-\bar{\nu}/2}\sum_\nu \frac{\bar{\nu}^{\nu/2}}{\sqrt{\nu!}}|\nu\rangle. \quad (7)$$

Following previous models, we also assume a constant vibration damping lifetime $\tau_k$, which doesn't change with the value of $\nu$. For a coherent state of vibration in mode k under a laser intensity of I, we have the following rate equation at equilibrium

$$e^{-\bar{\nu}}\sum_\nu \frac{\bar{\nu}^\nu}{\nu!}(\nu+1)r_{k,\nu}I = \bar{\nu}/\tau_k, \quad (8)$$

in which the left side is the total phonon pumping rate, which is a summation over all molecular vibration eigenstates, each item being the product of the probability to be in the state $|\nu\rangle$ according to equation (7) and the corresponding phonon pumping rate according to equation (6). The right side is the phonon damping rate. When all $r_{k,\nu}$ values are equal, equation (8) reduces to a single-step spontaneous and stimulated phonon emission problem of

$$(\bar{\nu}+1)r_{k,\nu}I = \bar{\nu}/\tau_k. \quad (9)$$

Assume $r_{k,\nu} = r_{k,2}$ for all values of $\nu \geq 2$, as mentioned earlier, equation (8) reduces to

$$e^{-\bar{\nu}}[(r_{k,0}-r_{k,2})+2\bar{\nu}(r_{k,1}-r_{k,2})]I + (\bar{\nu}+1)r_{k,2}I = \bar{\nu}/\tau_k. \quad (10)$$

The PERS signal power is proportional to $\bar{\nu}$ at equilibrium and given by

$$P_{SERS} / \hbar \omega_S = \bar{\nu} / \tau_k. \tag{11}$$

By equation (10), the relationship of average phonon number $\bar{\nu}$ versus laser intensity I is plotted in Figure 4 for different ratios between the $r_{k,\nu}$'s. As shown, when the $r_{k,\nu}$'s (or $|R_{k,\nu}|^2$'s) are equal, the phonon pumping behavior, plotted in blue, is the same as in the previous models, that is, a nonlinearity resulting from single-step stimulated phonon emission.[13,15] However, for a resonant Raman molecule whose $|\nu=0\rangle \to |\nu=1\rangle$ transition rate coefficient $r_{k,0}$ (or $|R_{k,0}|^2$) is fifteen times as large as that of the other transitions, the phonon pumping behavior is composed of two different steps, as plotted in green. First, under low laser power, the $|\nu=0\rangle \to |\nu=1\rangle$ transition dominates due to its large Raman activity, and it gradually saturates as the $|\nu=0\rangle$ state is being depleted. Then, as the laser power further increases, filling of higher phonon number states leads to stimulated phonon emission and a turn-on behavior of phonon pumping. Another situation in which the $|\nu=1\rangle \to |\nu=2\rangle$ transition is also enhanced, by a factor of five, is plotted in red. In this situation, the initial phonon pumping and saturation step involves one more state than in the second situation, and is saturated at a higher value of $\bar{\nu}$. Moreover, if the strongest transition starts with a different state than the $|\nu=0\rangle$ state, the saturation step can appear in the middle of the $\bar{\nu}$ versus I curve, which is not plotted.

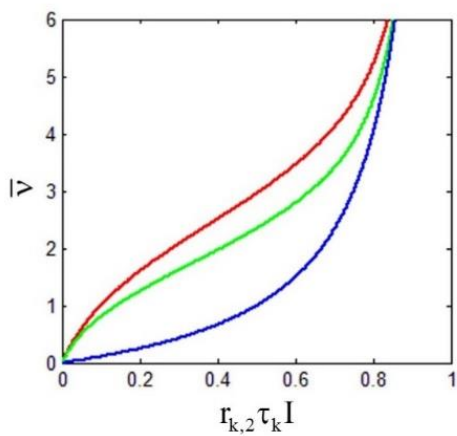

**Figure 4. Phonon pumping by Raman scattering.** $r_{k,0} : r_{k,1} : r_{k,2} = 1:1:1$ (blue), 15:1:1 (green), and 15:5:1 (red).

To conclude our theoretical work, we point out that the above model is still valid if the molecule has multiple orthogonal vibration modes, by summing over all the values of the subscript k. At last, it is important to point out the possibility that, in addition to Raman scattering, phonon pumping is also realized by the absorption and fluorescence process. Although MGITC is known as a non-fluorescent molecule, under extremely high enhancement, it is possible that Kasha's rule will break down and fluorescence will happen due to a strong Purcell effect.[19] However, the possible fluorescence from MGITC is buried in its strong background impurity emission, as evident for MGITC powders on glass substrates, which doesn't allow us to make a quantitative discussion on the effect of fluorescence on our experimental results here. An over-simplified model for the fluorescence effect is described in Supplementary Information.

**Experiment results.** Experimental observation of stepwise quantum phonon pumping shall be a piece of evidence for our theory as well as for the molecular cavity optomechanics model. An example of our measurement results is shown in Figure 5, in which the SERS intensity versus laser power relation is measured for the three Raman bands of MGITC from the same antenna, using the experiment described by Figure 1. A fitting to equation (10), except for the roll-over part, is also shown, which qualitatively agrees with the theoretical prediction of two-step phonon pumping. To confirm the validity of our observation, we measured another two antennas with laser power below the roll-over level and repeated each measurement for three times. These measurement results are presented in Fig. S2 of Supplementary Information, and clearly show good reproducibility and stability.

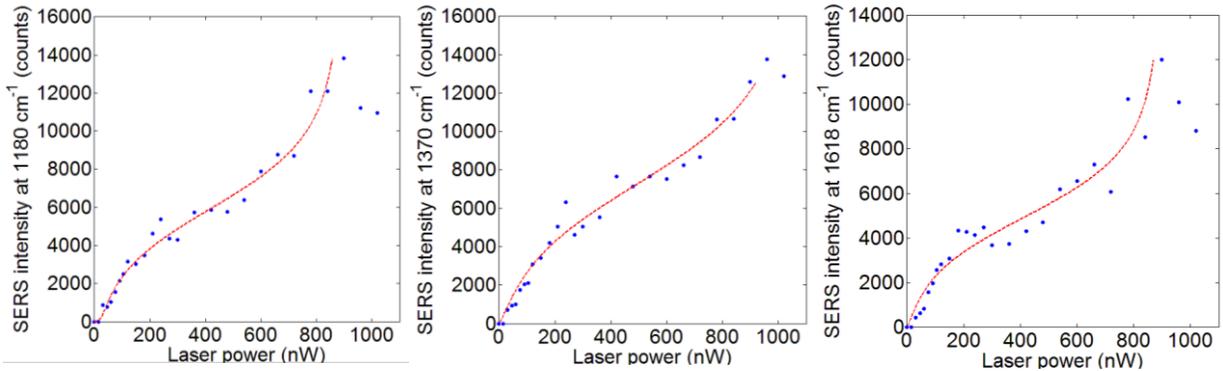

**Figure 5. The SERS intensities of three Raman bands of MGITC versus laser power at sample, for an antenna coated with a monolayer of MGITC.** The blue dots are measurement results. The red dashed curves are fitting to equation (10) except for the roll-over parts. The integration time is 4 s. Fitting parameters: $\nu = 0.00052/0.00044/0.00071 \times$ SERS intensity (counts), $r_0 : r_1 : r_2 = 11.1:11.1:1 \ / \ 12.2:12.2:1 \ / \ 16.7:16.7:1$, from left to right. The condition $r_0 \geq r_1 \geq r_2$ is enforced in the fitting just to reduce the fitting freedom but not due to any physical reasons.

## Discussion

(i) A natural question would be whether there are other possible mechanisms to reproduce the S-like shape of the PERS intensity versus laser power as in Figure 5. By looking at equation (6), photon stimulation and phonon stimulation are probably the only processes that may introduce nonlinearity, the former excluded by the low laser power in our experiment.

(ii) A numerical estimation on the laser power level required to reach $\bar{v} \geq 1$, and comparison with the experiments should be valuable. However, a simple calculation using empirical values of Raman scattering cross sections indicates that vibration damping lifetimes, $\tau_k$, that are orders of magnitude longer than what are commonly known for molecules in solutions must be used in the theory, for both our experiment and the previous work.[6,7,16]

First, let's estimate $\tau_k$ in our experiment. Since the exact Raman scattering cross section of the MGITC molecule, including the chemical enhancement factor, is unknown, a more accurate estimation is to use equation (11). In our experiment, each Raman photon emitted from the molecules in the hotspot will result in roughly 0.06 counts of readout by an electron multiplying charge coupled device (EMCCD) detector, as explained in Supplementary Information. Therefore, a $10^4$ count per 4 s Raman peak in our experiment with a spectral width of roughly 20 EMCCD pixels corresponds to $8\times10^5$ Raman photons per second scattered off all the molecules in the hotspot. Substituting this number into equation (11), in order for $\bar{v} \geq 1$, $\tau_k$ should be no less than 1 $\mu$s.

The value of $\tau_k$ is of central importance to many problems.[20] Vibration damping of polyatomic molecules in condensed phases has been attributed to intramolecular vibrational relaxation (IVR) by anharmonic coupling between different vibration modes, and intermolecular energy exchange and interaction with the low frequency continuum of the medium.[20,21] Time resolved vibrational spectroscopy experiments for molecules in liquids showed $\tau_k$ values on the picosecond level.[22,23] On the other hand, for molecules in the gas phase, microsecond level relaxation time is indeed possible, which is attributed to severe inhibition of energy flow between molecular vibration states of different polyad numbers.[24]

In our experiment, the samples are dry, and the remaining water molecules, if any, will be driven away from the ultra-intense hotspots. Therefore, the long $\tau_k$ values can be attributed to the lack of a liquid medium for the MGITC molecules to exchange energy with, and also the possible inhibition of IVR by the polyad bottleneck. For future work, time resolved measurement of PERS in a single dry hotspot shall reveal interesting information about molecular dynamics, the result of which will confirm or disprove the theory in this and previous work.

(iii) Both Raman linewidth narrowing by opto-mechanical feedback[13] and spectral shift as $v$ increases due to anharmonicity are expected in theory, but not resolved in our or previous experiments. They are claimed to be mitigated by linewidth broadening due to IVR and anharmonicity in ref. 13. In addition, here the Raman linewidth is from a collection of molecules each of which occupies approximately 1 nm$^2$ [25] in a 9 nm$^2$ hotspot[8], which is heterogeneously broadened.[26]

(iv) The PERS roll-over behavior in Figure 5 and ref. 8 is interesting since it happens at a much lower hotspot intensity than a previous report, which rules out temperature and photon absorption as the major photobleaching mechanisms. In 2005, Pettinger *et al.* characterized the photobleaching of MGITC on Au(111) surface in a TERS experiment.[27] Using a 0.5 mW He-Ne laser focused through a 50× microscope objective and directed to the sample at an angle of incidence of 60°, they reported a bleaching time constant of 6-7 s. Their $|\tilde{E}^2_{Loc}|$ =2500 and EMEF=6×10$^6$. In our experiment, roll-over happens near 1 $\mu$W, $|\tilde{E}_{Loc}(\omega_L)|^2$ is around 1×10$^4$ and EMEF is around 10$^{9.2}$.[8] Therefore, the photobleaching or roll-over hotspot intensity, evaluated by laser focal spot intensity×$|\tilde{E}^2_{Loc}|$, is over an order of magnitude weaker in our experiment than in the TERS experiment. This indicates that the molecules are not damaged by factors that are linear to hotspot intensity, such as photon absorption, temperature increase, *etc.* On the other hand, the molecular vibration energy, $\bar{v}\hbar\Omega_k$, which is determined by laser focal spot intensity×EMEF according to Figure 4, is about equal in both experiments. This is a hint that photobleaching in PERS with ultrahigh EFs could be the result of phonon pumping, that is, stimulated phonon emission efficiently converts laser power into molecular vibrations and shakes the molecules away from their initial positions and orientations.

(v) In addition to the fact that different antennas show different nonlinear fitting coefficients, 30% of the antennas measured don't even clearly show the two-step nonlinearity feature which are not included in this report. This indicates a complicated molecules-in-hotspot system which don't have reproducible intra-/inter-molecule interactions or coupling to the environment, where the molecules' position, orientation, chemical enhancement, resonance Raman enhancement, radiative and non-radiative processes, and coupling to medium could all have some effects. One possible reason is when different molecules reach the saturation step under different laser powers, the total Raman intensity behaves not as much S-shaped. For future work, numerical modeling of MGITC-Au's band structure, Raman activity and fluorescence at different positions and orientations in the hotspot, together with the experimental measurement of $\tau_k$, will improve the accuracy of our modeling and advance understanding about the variation in nonlinearity. Without these pieces of information, the fitting curves in Figure 5 are no more than an eye guide, and in fact a wide range of fitting parameters can be adopted to generate fitted curves close to the experimental data.

(vi) Measuring anti-Stokes scattering is also important for future work. It has been a critical method to study vibrational pumping, but only in the weak pumping regime, while stimulated phonon emission was predicted early in 2006.[15,28,29]

In conclusion, with the ultrahigh enhancement of Stokes Raman scattering in plasmonic hotspots, a considerable number of molecular vibration phonons can be produced at low laser powers, and we have theoretically modeled and experimentally demonstrated the stepwise quantum phonon pumping phenomenon in such a situation. In the theory, the molecular vibration is treated as a coherent state, while the Raman activities are set to be different for different phonon number states. In the experiment, the spectral alignment between the laser, the molecular energy level, the Stokes shift and the LSPR provides an ideal system to match our simple theoretical model, and the reproducible ultrahigh EFs of the nanosphere-plane antennas under RP excitation provide an efficient approach for experimental observation. The theory and experiments agree with each other to show a saturation step and a stimulated phonon emission step in the PERS intensity versus laser power relation. In addition, by comparing with previous experimental results on photobleaching, phonon pumping is suggested to be the cause of photobleaching under ultrahigh EFs. For future work, measurement of vibration damping time, modeling of probe molecule's band structure and transition rates, and study of anti-Stokes scattering will provide important new knowledge about the origin of PERS nonlinearity.

## Methods

Most of the following is selectively copied from the *Materials and Methods* part of ref. 8.

**Sample preparation.** First, 13.5 μL of 45 μM MGITC (Invitrogen M689) ethanol solution and 1 mL of $5.2 \times 10^9$/mL gold nanosphere ultra-purified water solution (BBI Solutions, 60 nm mean diameter, ± 8% variation) were incubated together for 2 hours at room temperature. Then the functionalized gold nanosphere solution was 1:1 diluted with ultra-purified water, and drop-casted onto the gold planes. The gold planes were 200 nm thick

Au(111) films on mica substrates (PHASIS), which had been deposited by magnetron sputtering and hydrogen flame annealed to obtain atomically flat surfaces. After 45 minutes, the samples were dried under a stream of nitrogen.

**Raman scattering measurement.** A He-Ne laser working at 632.8 nm and $TEM_{00}$ mode was used to excite the molecules. The laser beam passed through a liquid crystal polarization converter (ARCoptix) and was converted to the RP state. The RP laser beam was focused on to each antenna on the sample through a long working distance 100× Plan Apo objective, whose NA is 0.9. The laser power at sample was measured right after the focusing objective with a silicon photodiode. Raman scattering from the sample was collected by the same objective, passed through a long-pass filter, and detected by a monochromator installed with an EMCCD detector.

**LSPR spectra measurement.** A super continuum source was focused onto each antenna through the same 100× objective as in the Raman scattering measurement. The scattered light was collected outside the NA of the objective with a lens whose NA is 0.15. The collecting lens focused the scattered light into a fiber-bundle directed to the monochromator. The power of the super continuum source was carefully decreased by neutral density filters in order not to damage the samples.

**Identification of nanospheres.** The small scattering cross section of the antennas and the large reflection off the gold plane render it extremely difficult to find the nanospheres under optical microscopes without special methods. The same 100× objective was used as part of a home-built microscope to observe the nanospheres. A spatial filter blocked the central part of the objective's entrance pupil so that the nanospheres were illuminated at an inclined angle. The nanospheres appeared as dark spots on a bright background, due to the antennas' absorption and scattering of the inclined illumination. In addition, position markers were made by focused ion beam milling on the gold planes prior to nanosphere distributions, and scanning electron micrograph (SEM) images were taken before the SERS experiment to compare with the optical microscopy images, so that the nanospheres could be identified repeatedly. It is important to limit the SEM imaging dose upon the antennas so as not to damage them and to mitigate carbon deposition. Around 10% of the gold nanoparticles have irregular non-spherical shapes under SEM, and were excluded from measurement.

## Acknowledgements

This work is supported by the National Natural Science Foundation of China under grant # 11574207, the Science and Technology Commission of Shanghai Municipality under grant # 14JC1491700, and the Center for Advanced Electronic Materials and Devices of Shanghai Jiao Tong University. T.Y. thanks Christophe Galland from Ecole Polytechnique Fédérale de Lausanne (EPFL) for helpful discussions.

## Author contributions

T.Y. and J.L. designed the initial experiments and analyzed the data to discover the stepwise nonlinear behaviors. T.Y. and X.W. designed the further experiments and analyzed the data to confirm the reliability of measurement results. T.Y. built the theoretical model, and X.W. finished fitting the experimental data to the theory. J.L. and X.W. conducted the experiments. T.Y. wrote the paper with the other authors' assistance.

## Additional information

**Supplementary Information** accompanies this paper at http://www.nature.com/naturecommunications.

**Competing financial interests**: The authors declare no competing financial interests.

# Supplementary Information

# Stepwise quantum phonon pumping in plasmon-enhanced Raman scattering

Tian Yang, Jing Long and Xiaodan Wang

## 1. An over-simplified model for stepwise quantum phonon pumping of fluorescent molecules

Resonant Raman scattering is associated with optical absorption and fluorescence of the molecules. In such a situation, an important change to the model described in the main text is that, in addition to Raman scattering, phonon pumping is also realized by the absorption and fluorescence process. That is, the molecule, initially in state $|v\rangle$, absorbs a laser photon and spontaneously emits another photon, ending in state $|v'\rangle$. In addition to LSPR enhanced fluorescence, non-radiative coupling to surface plasmon polaritons (SPP) and lossy surface waves also contribute to phonon pumping, while other non-radiative processes can be ignored for a fluorophore with a high internal quantum efficiency. These extra phonon pumping pathways compete with the Stokes process and can significantly change the PERS intensity versus laser power relation curve.

In our experiment, the fluorescence lifetime is hugely shortened by the factor $|\tilde{E}_{Loc}^2|$ due to the Purcell effect, which allows us to make the following assumptions in addition to assumptions (1-5) in the main text: (6) The molecule is almost always in its electronic ground state under the low laser powers in our experiment. (7) When the molecule is in the electronic excited state, against the Kasha's rule, it doesn't have time to relax down the vibrational sub-structure before fluorescence happens [19]. Therefore, a significant portion of the fluorescence ends in the $|v+1\rangle$ state for the three Raman bands we pay attention to, since these fluorescence photons have their frequencies near the peak of LSPR. Such a $|v\rangle \to |v+1\rangle$ phonon pumping pathway is similar to the Stokes process, but the fluorescence rate is not phonon stimulated.

In the following, we give the equations by considering $|v\rangle \to |v+1\rangle$ absorption-fluorescence as the only phonon pumping pathway. Equation (8) becomes

$$e^{-\bar{v}} \sum_v \frac{\bar{v}^v}{v!} f_{k,v} I = \bar{v}/\tau_k, \tag{S1}$$

where $r_{k,v}$ is changed to $f_{k,v}$, the absorption-fluorescence rate coefficient. The PERS signal power is given by

$$P_{SERS}/\hbar\omega_S = e^{-\bar{v}} \sum_v \frac{\bar{v}^v}{v!} (v+1) r_{k,v} I. \tag{S2}$$

When all $f_{k,v}$ and $r_{k,v}$ values are equal, Equations (S1) and (S2) reduce to a linear phonon pumping problem,

$$f_{k,v} I = \bar{v}/\tau_k. \tag{S3}$$

Its PERS intensity versus laser power relation follows a second order polynomial,

$$P_{SERS}/\hbar\omega_S = (\bar{v}+1) r_{k,v} I = r_{k,v}(f_{k,v}\tau_k I^2 + I), \tag{S4}$$

which is another source of nonlinearity for experimental observations, and which becomes more important when Raman scattering is not enhanced enough to compete with fluorescence. Assume $f_{k,v} = f_{k,2}$ and $r_{k,v} = r_{k,2}$ for all values of $v \geq 2$, Equations (S1) and (S2) reduces to

$$e^{-\bar{v}}[(f_{k,0} - f_{k,2}) + \bar{v}(f_{k,1} - f_{k,2})]I + f_{k,2} I = \bar{v}/\tau_k, \tag{S5}$$

$$P_{SERS}/\hbar\omega_S = e^{-\bar{v}}[(r_{k,0} - r_{k,2}) + 2\bar{v}(r_{k,1} - r_{k,2})]I + (\bar{v}+1) r_{k,2} I. \tag{S6}$$

A numerical example of the relations of average phonon number, PERS and fluorescence intensities versus laser intensity is plotted in Figure S1, considering both phonon pumping pathways, that is, Raman scattering and $|v\rangle \to |v+1\rangle$ absorption-fluorescence. For comparison, the situation where fluorescence is absent is also plotted. As shown, under low laser intensities, fluorescence lowers the Raman scattering intensity by depleting the initial $|v\rangle$ state. Under higher laser intensities, fluorescence increases the number of phonons and leads to stronger phonon stimulated Raman scattering.

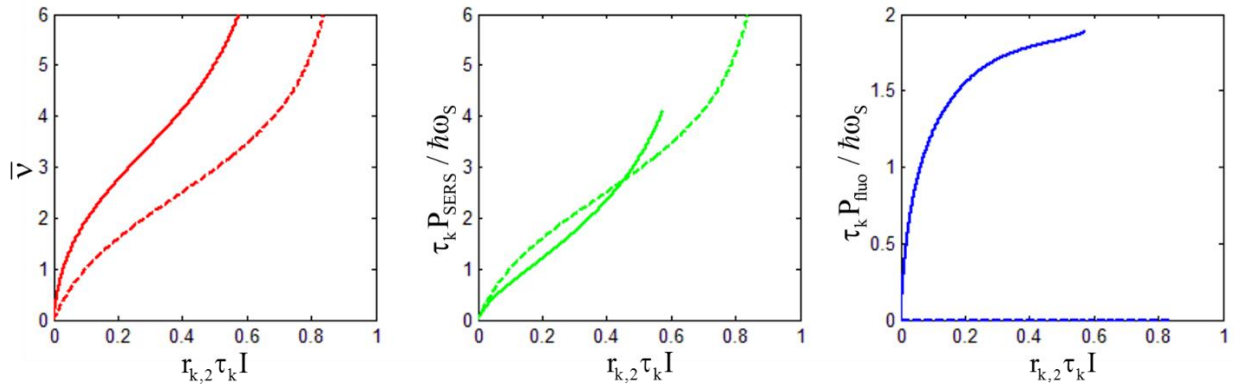

**Figure. S1**. **Phonon pumping by Raman scattering and fluorescence.** $r_{k,0}:r_{k,1}:r_{k,2}:f_{k,0}:f_{k,1}:f_{k,2} = 15:5:1:45:15:3$ (solid), 15:1:1:0:0:0 (dashed).

It should be kept in mind that the above model for fluorescent molecules could be oversimplified. For example, electronic down transitions through coupling to SPPs and lossy surface waves with shorter wavelengths than the laser will cool down the molecular vibration instead of increasing the vibration phonon number, by which Raman scattering under low laser intensities will be increased due to less depletion of the lower $|v\rangle$ states which correspond to resonance enhanced Raman scattering, and phonon stimulated Raman scattering under higher laser intensities will be decreased due to a smaller number of phonons.

In the experiment, when the SERS intensity versus laser power relation shows a significant nonlinearity, often so does the background intensity of the SERS spectrum. However, without LSPR enhancement, MGITC molecules show no sign of fluorescence but a weak light emission background from impurities. Therefore, it is difficult to identify the source of the SERS spectrum background in our experiment at the moment, and a discussion on the effect of fluorescence from the experimental aspect will not be conducted in this paper.

**2. More experimental results on MGITC SERS intensity versus laser power relation**

In Figure S2, we show the measurement results of SERS intensity versus laser power relation for another two antennas each coated with a monolayer of MGITC molecules. The measurement for each antenna has been repeated for three times, each time from low laser power to high laser power but below the roll-over level (the laser power was increased until the first roll-over point appears, which was not plotted). The results show a good reproducibility and stability of our measurement, and present convincing evidence for the claimed nonlinearity.

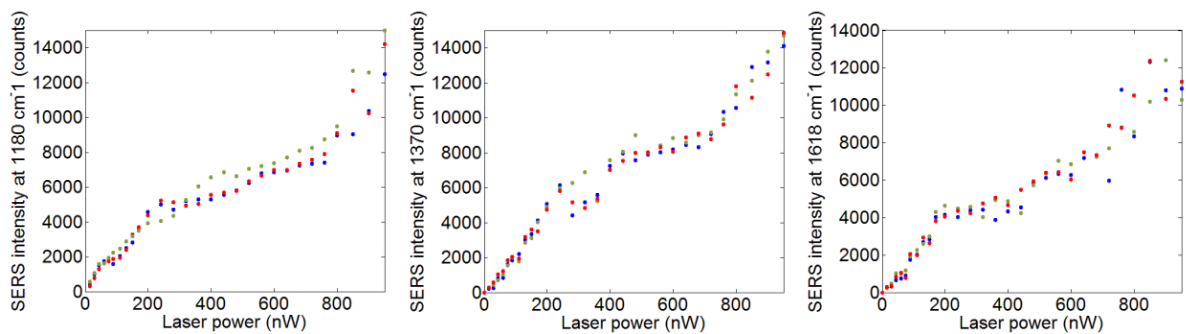

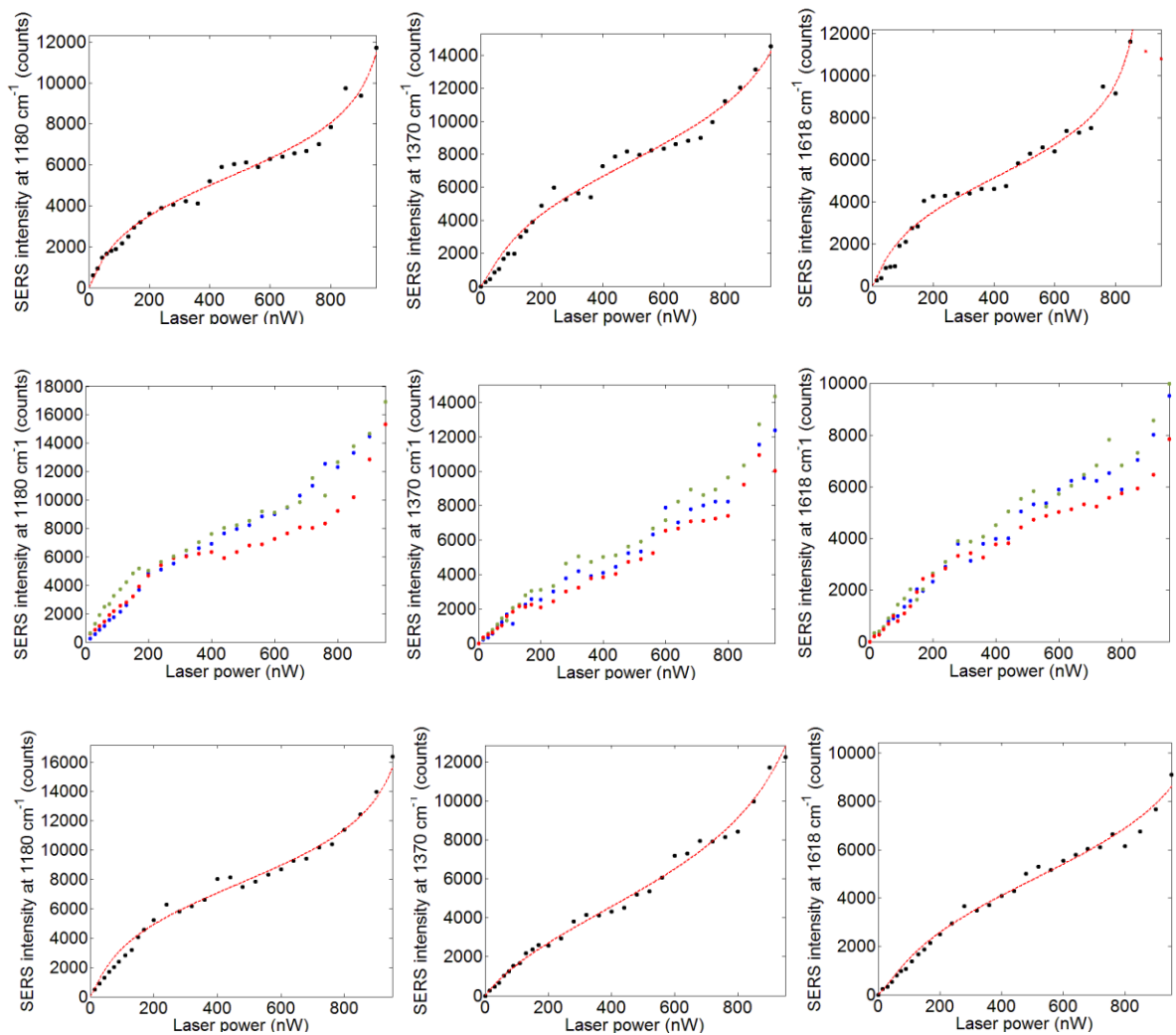

**Figure. S2. The SERS intensities of three Raman bands of MGITC versus laser power at sample, for another two antennas coated with a monolayer of MGITC.** The first and second rows correspond to the first antenna, and the third and fourth rows correspond to the second antenna. In the first and third rows, the colored dots represent three measurements, respectively, each from low laser power to high laser power but below the roll-over level, in the order of blue, red then green. In the second and fourth tows, the black dots are the average SERS intensities of the three measurements. The red dashed curves are fitting to Equation (10) except for the roll-over parts. The integration time is 4 s. Fitting parameters from left to right: second row: $\nu = 0.00066/0.00041/0.00065 \times$ SERS intensity (counts), $r_0 : r_1 : r_2 = 16.8:16.8:1/10.0:10.0:1/13.68:13.68:1$; fourth row: $\nu = 0.00045/0.00029/0.00059 \times$ SERS intensity (counts), $r_0 : r_1 : r_2 = 15.6:15.6:1/6.75:6.75:1/7.76:7.76:1$.

## 3. Estimation of Raman scattering detection efficiency

The Raman scattering detection efficiency is roughly estimated to be 0.06 counts/photon. In this estimation, we have assumed that 100% of the Raman photons scattered off the molecules are coupled to LSPR of the nanosphere-plane antenna; 22% of the vertical antenna LSPR power is radiated into far-field, with the rest coupled to ohmic loss, lossy surface waves and SPPs; the antenna far-field radiation is collected by a NA=0.9 objective with a collection efficiency of 59%; then the photons undergo power losses due to transmission through lenses, a beam splitter, multiple mirror reflections, an optical filter and spectrometer grating diffraction, with a total optical transmission of 5.6%; finally the EMCCD produces 100 counts per 5.7 electrons with a 50% quantum efficiency.